\documentstyle[preprint,aps,eqsecnum]{revtex}
\begin{document}

\thispagestyle{empty}
\setcounter{page}{1}

\title{MAGNETIC SCREENING AT FINITE TEMPERATURE}

\author{Cristina Manuel}

\address{Dpt. Estructura i Constituents de la Mat\`{e}ria\\
Facultat de F\'{\i}sica,
Universitat de Barcelona \\
Diagonal 647,
08028 Barcelona (SPAIN)} 

\maketitle

\thispagestyle{empty}
\setcounter{page}{0}

\begin{abstract}
$\!\!$It is shown that at finite temperature and in the presence of
magnetic sources magnetic fields are screened.
This is proven within the framework of
classical transport theory both for the
Abelian and non-Abelian plasmas. Magnetic screening arises in this
formalism as a consequence of polarization effects occurring in the
plasmas, and it is proportional to the inverse of the gauge coupling constant.
It is then discussed whether this mechanism could be relevant in 
realistic quantum gauge field theories, such as QCD.
\end{abstract}

\vfill

\hfill\break
\hbox to \hsize{ECM-UB-PF-96/24} 
\hbox to \hsize{December/1996}
\vskip-12pt
\eject

\baselineskip= 20pt
\pagestyle{plain}
\baselineskip= 20pt
\pagestyle{plain}

\baselineskip=15pt
\pagestyle{plain}

\section{INTRODUCTION}
\label{Intro}

The studies of gauge field theories at high temperature $T$ have attracted
much attention in the recent past \cite{LeBellac}. 
 The behavior of certain gauge theories, such as QCD, changes dramatically
in their  low or high  temperature regimes. Thus, it is generally believed
that QCD is in a  unconfined phase above a critical temperature $T_c \sim 200$ 
MeV. In principle, due to asymptotic freedom, 
perturbation theory in the high $T$ regime of QCD could be
naively expected to be valid, but this turns out not to be the case.

In the high $T$ regime of QCD, 
thermal excitations produce a plasma of charged particles which screens
color electric fields. The electric thermal mass can be computed at
one-loop order in perturbation theory, and it is proportional to $g T$,
where $g$ is the gauge coupling constant. A self-consistent inclusion
of the color electric mass in the Feynman loop
computations  requires the use of resummed
perturbation theory \cite{BP}.
 
Color magnetic fields are not screened at the same order of perturbation
theory. Due to this fact several infrared (IR) divergences are encountered
in the computations of different physical quantities. Those divergences
even appear in the absence of quark matter in the computation of 
the perturbative partition function \cite{Gross}.

There has been a lot of discussions in the literature
about the possible mechanism that
could cure the IR problems in the magnetic sector of a pure Yang-Mills
theory. It is generally believed \cite{Gross} that non-perturbative effects
generate a color magnetic mass of order $g^2 T$, which would 
arise  only in the non-Abelian theory but not in the Abelian case. The arguments
to reach to that conclusion are the following. The infrared limit of QCD
 at high $T$ is governed by the spatial vector
 gauge field $A_i$. Using imaginary time formalism, the high $T$ limit of 
QCD is equivalent to an Euclidean three dimensional Yang-Mills theory,
with an effective coupling constant $g^{'2}= g^2 T$. Those theories are
believed to generate a dynamical mass gap proportional to the (dimensional)
coupling constant, which is not computable in perturbation theory.

The above arguments are rather qualitative, but give no clue about what
kind of mechanism could be responsible for the generation of the magnetic
mass gap. They also do not explain how those effects could eliminate the
IR problems found in specific computations.

Some attempts have been made in the literature to give an effective action that
describes a magnetic mass term and such that could be used in specific
Feynman loop diagrams computations \cite{Nair}. However, 
no conclusive evidence about the proposed {\it Ans\"{a}tze}  has been reached.

On the other hand, some discussions have been raised in the literature
about whether the high $T$ limit of QCD could be correctly described 
within the  classic rather than quantum field theory. In Ref. \cite{Ambjorn}
it has been realized that both the color electric and magnetic screenings are 
detected in lattice  classical gauge field computations.
This fact could be a clear indication that the magnetic screening could
be understood in terms of classical but non-perturbative physics.

The purpose of this article is to investigate  under which circumstances
static magnetic screening can emerge in a hot non-Abelian theory
using exclusively classical (or semiclassical) physics. I do not
attempt here to solve the infrared problem of a Yang-Mills theory
at finite temperature, but I just try to find a classical mechanism that
could generate a thermal color magnetic mass.

A study of the classical gauge field equations reveals that  
the absence of color magnetic screening in a hot plasma is entirely due
to the absence of static magnetic sources.    Therefore, in order to derive
a thermal magnetic screening,  I
{\it postulate} the existence of magnetic charges. It will be  shown in 
this article that in the presence of magnetic charges  magnetic fields are 
naturally screened at finite $T$.
The mechanism responsible for this screening
is actually  the same one as that generating the screening
of electric fields in the QED and QCD plasmas.
In  hot plasmas of electric charges, polarization
phenomena  screens  electric fields. The
same kind of polarization effects would  generate the magnetic
screening if magnetic charges  existed.

Classical transport theory will be used to 
prove that the existence of (non-) Abelian magnetic charges 
at finite $T$ implies the screening of (non-) Abelian magnetic fields.
This  formalism  has already been used to 
derive the screening of (non-) Abelian electric fields, and those effects
are reproduced exactly in the corresponding quantum field theory at 
high $T$.  The  generalized set of (non-) Abelian Vlasov equations
in the presence of magnetic sources will be written, and from them 
magnetic screening will be derived. As it will be shown, the
magnetic screening turns out to be  proportional to the inverse 
gauge coupling constant.

This paper is structured as follows. In Sec.\,\ref{Abelian}, 
the Abelian plasmas are first studied.
 In Subsec.\,\ref{AbDebye}  it is recalled how
Debye screening is obtained from the Vlasov equations.
In Subsec.\,\ref{DAbDebye} the 
proposed ``magnetic" or dual Vlasov equations are written, and from them
the static screening of magnetic fields is derived.
 It is stressed there that the duality symmetry of electromagnetism allows
to derive the magnetic screening from the electric one.
In Sec.\,\ref{Nonabelian}  the same study is reproduced for
non-Abelian plasmas, when duality is not a symmetry of the theory.
Subsec.\,\ref{Vacsmon} is devoted to review  some  static
magnetic monopole fields  solutions in the vacuum.
In Subsec.\,\ref{Hotsmon}
solutions to the gauge field equations in  hot plasmas, which reproduce
the screened magnetic fields, are found.
Sec.\,\ref{conclu} ends  with a discussion
of the results. Let us finally mention that
throughout this paper a system of units where $\hbar = c = k_B = 1$ will be
used.

\section{THE ULTRARELATIVISTIC ABELIAN PLASMAS}
\label{Abelian}

\subsection{Static Electric Screening in the Plasma of Electric Charges}
\label{AbDebye}

In this subsection the derivation of the electric Debye screening effects
from classical kinetic equations is reviewed
\cite{Landau}, \cite{Silin}. This will teach us how to
derive the corresponding magnetic screening in the presence of
magnetic charges.

Let us first recall the dynamical
evolution of a charged point particle.
A particle carrying an electric charge $e$, with mass $m$, and transversing
a worldline $y^{\mu} (\tau)$, where $\tau$ is the proper time,         
obeys the equations (neglecting the effects of spin)
\begin{mathletters}
\label{lorentz}
\begin{eqnarray}
m\, {{d  y^{\mu}(\tau) }\over{d \tau}} & = &  p^{\mu} (\tau) 
\ , \label{lora} \\[2mm]
m\, {{d  p^{\mu} (\tau)}\over{d \tau}} & = &
 e\,  F^{\mu\nu} ( y(\tau)) \,  p_{\nu} (\tau)
\ ,\label{lorb}
\end{eqnarray}
\end{mathletters}
$\!\!$where the electromagnetic field $F^{\mu\nu}$ 
is evaluated on the particle worldline.
In our conventions $F ^{0 i}= - E^i$, and  $F ^{i j}= - \epsilon^{ijk} B^k$,
where $E^i$ and $B^i$ are the electric and magnetic fields, respectively.

In a self-consistent picture the electromagnetic
fields obey  the Maxwell's equations which have as sources the electric
currents obtained from each charged  particle of the system. Thus
\begin{equation}
\partial_\nu F^{\nu\mu} (x) = J^{\mu}(x) =  \sum_{\rm species} \, \sum_{\rm
helicities} \,
j^{\mu}(x)\ ,
\label{Maxwell}
\end{equation}
where (helicity and species indexes are implicit)
\begin{equation}
j^{\mu}(x) = e  \int d \tau \,  { d y^{\mu}(\tau) \over d \tau} \,   
\, \delta^{(4)} ( x - y (\tau)) \ . 
\end{equation}
The above current is conserved, $\partial_{\mu} j^{\mu}(x) = 0$, as 
may be checked by using the equations of motion (\ref{lorentz}).
This is required as a compatibility condition, as can be easily recognized by 
applying a partial derivative $\partial_{\mu}$ to (\ref{Maxwell}), since $F^{\mu\nu}$ 
is antisymmetric in their indices. 

The electromagnetic field  tensor obeys the Bianchi identity
\begin{equation}
\partial_\nu \, ^*F^{\nu\mu} (x) = 0 \ ,
\label{Bianchi}
\end{equation}
where the dual field is
 $^*F^{\nu\mu} = \frac12 \epsilon^{\nu\mu\rho\sigma} F_{\rho\sigma}$,
and $ \epsilon^{\nu\mu\rho\sigma}$ is the Levi-Civita antisymmetric tensor
in four dimensions, with  $ \epsilon^{0123} =1$.  In our conventions 
$^*F ^{0 i}= - B^i$, and  $^*F ^{i j}=  \epsilon^{ijk} E^k$. The Bianchi identity
(\ref{Bianchi}) guarantees the existence of the vector gauge field $A_\mu (x)$,
such that $F_{\mu\nu} = \partial_{\mu} A_{\nu} - \partial_{\nu} A_{\mu}$, which
is not unique. The ambiguity in defining the vector gauge field is that
corresponding to gauge transformations.

The statistical description of the plasma of charged particles
is given by the distribution function of its components in their phase-space.
In the collisionless case the one-particle distribution function $f(x,p)$ 
of finding a particle in the state $(x,p)$
evolves in time via a transport equation
\begin{equation}
\frac{d f(x,p)} {d \tau} = 0 \ .
\end{equation}
Using the equations of motion (\ref{lorentz}), it becomes the Boltzmann
equation
\begin{equation}
p^{\mu}\left[{{\partial}\over{\partial x^{\mu}}}
- e \,F_{\mu\nu} (x) {{\partial}\over{\partial p_{\nu}}}
\right] f(x,p) = 0 \ .
\label{boltzmannab}
\end{equation}
In a self-consistent picture, the mean electromagnetic fields obey
the Maxwell's equations
\begin{eqnarray}
\partial_\nu F^{\nu\mu} (x)  & =  &J^{\mu}(x) =  \sum_{\rm species}\
\sum_{\rm helicities}
j^{\mu}(x)\ , \\
\partial_\nu \, ^*F^{\nu\mu} (x) & = & 0 \ ,
\label{Maxwellt}
\end{eqnarray}
where now each particle species electric current is obtained from
the corresponding distribution function as
\begin{equation}
j^{\mu} (x) = e \int dP \, p^{\mu} \, f(x,p) \ .
\label{Bacurrent}
\end{equation}
 The momentum measure in (\ref{Bacurrent}) is defined as
\begin{equation}
dP = {{d^{4}p}\over{(2\pi)^{3}}}\,\,2\,\theta(p_{0})\,\,
\delta(p^{2} - m^{2}) \ , 
\label{measurep}
\end{equation}
so that it guarantees positivity of the energy and on-shell evolution.

The equations (\ref{boltzmannab})-(\ref{Bacurrent}) 
are known as  Vlasov equations \cite{Landau}.

The effects of static screening in the plasma of electrons and ions 
can be deduced from the Vlasov equations as follows \cite{Landau},
 \cite{Silin}. Let us consider
a neutral plasma, that is, composed by the same number of positive
and negative charges. The plasma, initially at equilibrium, is disturbed by
a weak electromagnetic field.
We look for a distribution function of the form
\begin{equation}
f(x,p) = f^{(0)} (p_0) + e \,  f^{(1)} (x,p) + ... \ , 
\end{equation}
where $  f^{(0)} (p_0)$ is, up to a normalization constant, the
Fermi-Dirac equilibrium function
\begin{equation}
n_F (p_0) = \frac{1}{e^{p_0/T} + 1} \ .
\end{equation}

Neglecting second order terms, $f^{(1)}$ obeys the following equation
\begin{equation}
p^{\mu} {{\partial}\over{\partial x^{\mu}}} f^{(1)} (x,p) =
p^{\mu} \,F_{\mu 0} (x) { d \over d p_{0}}  f^{(0)} (p_0)  \ .
\label{pertbolt}
\end{equation}
Notice that {\it only} the electric field enters into the r.h.s. of  (\ref{pertbolt}),
but not the magnetic field.
This is actually the reason why there is only static electric screening
but not static magnetic screening in this approach.

 A  total electric current density  $J^{\mu} (x, p)$ is defined such that the 
total current $J^{\mu} (x)$  is found just by integrating over the momenta
$p$, using the momenta measure (\ref{measurep}). 
The induced electric current density $J^{\mu} (x, p)$ obeys
the equation
\begin{equation}
p \cdot \partial \, J^{\mu} (x,p) =
 e^2 p^{\mu} p^{\nu} \,F_{\nu 0} (x) { d \over d p_{0}}  f^{(0)} (p_0)  \ .
\end{equation}

In the ultrarelativistic limit, that is, taking the fermion mass $m=0$ in
(\ref{measurep}), 
the induced current in momentum space
reads
\begin{equation}
J^{\mu} (k) = - i m^2 _D \int \frac{d \Omega_{\bf {\hat v}}}{4 \pi} \,
\frac{v^{\mu}}{v \cdot k + i \epsilon} \, {\bf v} \cdot {\bf E} (k) \ ,
\label{indcurmom}
\end{equation}
where $m_D ^2 = e^2 T^2 /3$ is the Debye mass squared. Retarded 
boundary conditions have been imposed in (\ref{indcurmom}),
with the prescription $i \epsilon$. The four vector
 $v^{\mu} = (1, {\bf {\hat v}})$ is the four velocity of the particles of the 
plasma, and in the ultrarelativistic situation considered here, it is 
light-like. The angular integral in (\ref{indcurmom}) is defined over
all possible directions of the three dimensional unit vector
 ${\bf {\hat v}}$. 

In the static situation  $J^{i} = 0$, while the induced
electric density  is
\begin{equation}
J^{0} ( {\bf x}) = - i m^2_D \int \frac{ d^3 k}{(2 \pi)^3} \,
\frac{ {\bf k} \cdot {\bf E} ({\bf k})}{{\bf k}^2} e^{ i {\bf k} \cdot
 {\bf x}}
\ .
\end{equation}

The Maxwell's equations which have as source
the current (\ref{indcurmom})   are also known as Kubo equations.
In the static limit they read
\begin{mathletters}
\label{Kuboabel}
\begin{eqnarray}
{\bf \nabla} \cdot {\bf E} \, ({\bf x}) & = & J^{0} ({\bf x}) \ , \qquad
{\bf \nabla} \times {\bf B} \, ({\bf x})  =  0 \ , \\
{\bf \nabla} \cdot {\bf B} \, ({\bf x}) & = & 0 \ ,  \qquad \qquad
{\bf \nabla} \times {\bf E}\, ({\bf x})  =  0 \ .
\end{eqnarray}
\end{mathletters}

These equations describe the static screening of electric fields
inside the plasma.
In the static situation the magnetic fields obey the same equations
as in the vacuum. In the non-static situation, magnetic 
fields also  suffer a dynamical screening in the plasma,
but we will not be concerned 
in this article about dynamical effects.

Let us finally recall that the static screening effects 
described in this subsection have been reproduced in the context
of  perturbative  QED in the high temperature limit.

\subsection{Static Magnetic Screening in the Plasma of Magnetic
Charges}

\label{DAbDebye}

In this subsection the effects of static screening in a
plasma of magnetic charges are derived.

Let us first recall the classical equations of motion of a particle
carrying a magnetic charge ${\tilde e}$, with mass $m$, and transversing
a worldline $y^{\mu} (\tau)$ \cite{WuYang1}
\begin{mathletters}
\label{dlorentz}
\begin{eqnarray}
m\, {{d  y^{\mu}(\tau) }\over{d \tau}} & = &  p^{\mu} (\tau) 
\ , \label{dlora} \\[2mm]
m\, {{d  p^{\mu} (\tau)}\over{d \tau}} & = &
{\tilde e}\,  ^*F^{\mu\nu} ( y(\tau)) \,  p_{\nu} (\tau)
\ .\label{dlorb}
\end{eqnarray}
\end{mathletters}
$\!\!$ In a self-consistent picture, these are augmented
with the field equations
\begin{eqnarray}
\partial_\nu F^{\nu\mu} (x)  & =  & 0 \ , \\
\partial_\nu \, ^*F^{\nu\mu} (x) & = & {\tilde J}^{\mu}(x) =  \sum_{\rm species} \,
\sum_{\rm helicities} \,
{\tilde j}^{\mu}(x)\ .
\label{dMaxwellt}
\end{eqnarray}
The magnetic current is computed for each particle species as
\begin{equation}
{\tilde j}^{\mu}(x) ={\tilde e}  \int d \tau \,  { d y^{\mu}(\tau) \over d \tau} \,   
\, \delta^{(4)} ( x - y (\tau)) \ ,
\label{magcurre}
\end{equation}
and it is conserved $\partial_{\mu} {\tilde j}^{\mu}(x) =0$.

Comparing Eqs.\,(\ref{lorentz})-(\ref{Bianchi}) and
(\ref{dlorentz})-(\ref{magcurre}), we see that they are symmetric
under the  interchange of electric and magnetic fields
$({\bf E}, {\bf B}) \rightarrow   ({\bf B}, -{\bf E })$,
and electric charges by magnetic ones. This is the so 
called duality symmetry of electromagnetism.

In the presence of a magnetic charge, and due to 
 the absence of the Bianchi identity (\ref{Bianchi}), it is not 
ensured that the electromagnetic field can be derived globally from a 
vector gauge field as $F_{\mu \nu} = \partial_{\mu} A_{\nu} -   
\partial_{\nu} A_{\mu}$. However, one can still define 
a vector gauge field which obeys that condition outside
the worldline of the magnetic charge. Therefore, it is possible
to define a vector gauge field $A_{\mu}$ locally. For the case of
the point magnetic charge, it is enough to define two different
vector gauge fields $A_{\mu}$ in different space-time regions, the two
solutions being related in their common domain of definition by a
gauge transformation\footnote{Alternatively, one could work with an
unique vector gauge field  $A_{\mu}$ with a Dirac string attached to 
the monopole. We prefer the Wu-Yang \cite{WuYang2} construction which
eliminates references to the Dirac string (see Subsec. \ref{Vacsmon}).}. However,
it would be possible to derive the dual electromagnetic field from a dual 
vector gauge field ${\tilde A}_{\nu}$, such that $^*F_{\mu \nu} = 
\partial_{\mu} {\tilde A}_{\nu} - \partial_{\nu}
 {\tilde A}_{\mu}$, since $\partial^{\mu} F_{\mu \nu} =0$, and 
$F^{\mu \nu} = - \frac 12 \epsilon ^{\mu \nu \alpha \beta} \,^*F_{\alpha \beta}$.

The above equations show the exact duality of electromagnetism
when electric and magnetic degrees of freedom are interchanged.
Due to this fact, one naturally expects that the propagation properties
of electric fields in the plasma of electric charges are
the same as those of magnetic fields in the plasma of magnetic
charges. In particular, this is true for  static screening.
Although from the duality symmetry arguments this is obvious,
let us show how one would derive the static screening
of magnetic fields.

Let ${\tilde f} (x,p)$ be the probability distribution 
function of finding a particle with mass $m$ carrying a magnetic charge
${\tilde e}$ in the state $(x,p)$. 
The corresponding set of dual Vlasov equations 
can be derived from the equations of motion (\ref{dlorentz}), and it
reads
\begin{eqnarray}
& p^{\mu}& \left[{{\partial}\over{\partial x^{\mu}}}
- {\tilde e} \, ^*F_{\mu\nu} (x) {{\partial}\over{\partial p_{\nu}}}
\right] {\tilde f}(x,p)  =  0 \ , \\
& \partial_\nu & F^{\nu\mu} (x)   =   0 \ , \\
&\partial_\nu &\, ^*F^{\nu\mu} (x)  =   {\tilde J}^{\mu}(x) = 
 \sum_{\rm species} \, \sum_{\rm helicities} \, {\tilde j}^{\mu}(x) \ .
\end{eqnarray}
$\!\!$  Now each particle species magnetic current is obtained from
the corresponding distribution function as
\begin{equation}
{\tilde j}^{\mu} (x) = {\tilde e} \int dP \, p^{\mu} \, {\tilde f}(x,p) \ .
\end{equation}

Let us consider a  neutral plasma of magnetic charges which is initially
at equilibrium. If the system  is perturbed by a 
weak field, then ${\tilde f}$ can be found in the form
\begin{equation}
{\tilde f}(x,p) = {\tilde f}^{(0)} (p_0) +  {\tilde e} \,  {\tilde f}^{(1)} (x,p) + ... \ , 
\label{expabm}
\end{equation}
where ${\tilde f}^{(0)}$ is, up to a normalization constant, 
 the  Fermi-Dirac equilibrium distribution function $n_F (p_0)$.
The equation obeyed by $ {\tilde f}^{(1)}$ reads
\begin{equation}
p^{\mu} {{\partial}\over{\partial x^{\mu}}} {\tilde f}^{(1)} (x,p) =
p^{\mu} \, ^*F_{\mu 0}(x){ d \over d p_{0}}  {\tilde f}^{(0)} (p_0)  \ .
\label{dualpertbolt}
\end{equation}
Notice that {\it only} the magnetic field enters into the r.h.s. of  (\ref{dualpertbolt}),
since $^*F_{i 0} = - B_i$.

In the ultrarelativistic limit, {\it i.e.} taking $m = 0$,
the conserved magnetic current in Fourier space is given by
\begin{equation}
{\tilde J}^{\mu} (k) = -  i {\tilde m}^2 _D \int \frac{d \Omega_{\bf {\hat v}}} {4 \pi} \,
\frac{v^{\mu}}{v \cdot k + i \epsilon} \, {\bf v} \cdot {\bf B} (k) \ ,
\label{dualindcurmom}
\end{equation}
where ${\tilde m}_D ^2 = {\tilde e}^2 T^2 /3$ is the magnetic Debye mass
squared.  The notation that  has been used above is the same as
in Eq. (\ref{indcurmom}), that is,  retarded boundary
conditions have been implemented,
 and $v^{\mu}$ is the four light-like velocity of the particle.

In the static situation ${\tilde J}^{i} = 0$, and the dual Kubo 
equations read 
\begin{mathletters}
\label{DKuboabel}
\begin{eqnarray}
{\bf \nabla} \cdot {\bf E} \, ({\bf x}) & = & 0 \ , \qquad \qquad
{\bf \nabla} \times {\bf B}\,  ({\bf x})  =  0 \ , \\
{\bf \nabla} \cdot {\bf B} \, ({\bf x}) & = &  {\tilde J}^{0} ({\bf x}) \ ,  \qquad 
{\bf \nabla} \times {\bf E}\, ({\bf x})  =  0 \ ,
\end{eqnarray}
\end{mathletters}
where 
\begin{equation}
{\tilde J}^{0} ( {\bf x}) = - i {\tilde m}^2_D \int \frac{ d^3 k}{(2 \pi)^3} \,
\frac{ {\bf k} \cdot {\bf B} ({\bf k})}{{\bf k}^2} e^{ i {\bf k} \cdot
 {\bf x}}
\ .
\end{equation}

These equations describe the static screening of magnetic fields. 
Notice that they could have been obtained 
 from  Eqs. (\ref{Kuboabel}) just by using
duality symmetry arguments.

Since the plasma is composed  of quantum particles,
the magnetic 
charge ${\tilde e}$ is subject to the Dirac quantization condition
\begin{equation}
e \, {\tilde e} = 2 \pi n \ , \qquad  n \, \epsilon  \, Z \ .
\end{equation}

Therefore, the magnetic Debye mass can be written in terms of the
electric charge as
\begin{equation}
{\tilde m}_D = \frac {T}{\sqrt{3} \, e} \, 2 \pi n  \ .
\end{equation}

Let us notice that in the perturbative regime $e \ll 1$, the expansion 
performed in Eq.(\ref{expabm}) would have failed.

\section{THE ULTRARELATIVISTIC NON-ABELIAN 
PLASMA}
\label{Nonabelian}

\subsection{Static Screening of Color Electric Fields in the Plasma
of Non-Abelian Charges}
\label{NAbDebye}

Consider a particle bearing a non-Abelian $SU(N)$
color charge $Q^{a}, \ a=1,...,N^2-1$, 
traversing a worldline $y^{\mu}(\tau)$.  The Wong equations 
\cite{Wong} describe the dynamical evolution of the 
variables 
 $x^{\mu}$, $p^{\mu}$ and 
$Q^{a}$ (we also neglect here the effect of  spin):
\begin{mathletters}
\label{wongeq}
\begin{eqnarray}
m\, {{d  y^{\mu}(\tau) }\over{d \tau}} & = &  p^{\mu} (\tau) 
\ , \label{wongeqa} \\[2mm]
m\, {{d  p^{\mu} (\tau)}\over{d \tau}} & = &
 g\,  Q^{a} (\tau)\, F^{\mu\nu}_{a} ( y(\tau)) \,  p_{\nu} (\tau)
\ ,\label{wongeqb}\\[2mm]
m\, {{d  Q^{a} (\tau) }\over{d \tau}} & = & - g\, 
f^{abc} \,  p^{\mu} (\tau) \,
A^{b}_{\mu} ( y(\tau)) \,  Q^{c} (\tau)\ , \label{wongeqc}
\end{eqnarray}
\end{mathletters}
$\!\!$where $f^{abc}$ are the structure constants of the group, 
$F^a _{\mu\nu}$ denotes the field strength, which is defined as
\begin{equation}
F^a _{\mu\nu} =  \partial_{\mu} A^a _{\nu} - \partial_{\nu} A^a _{\mu}
+ g f^{abc} A^b_{\mu} \, A^c  _{\nu} \ .
\label{defcol}
\end{equation}

The color fields obey the Yang-Mills equation
\begin{equation}
[D_\nu F^{\nu\mu}]^a(x) = J^{\mu\, a}(x) =  \sum_{\rm species} \,
 \sum_{\rm helicities}\ 
j^{\mu\, a}(x)\ ,
\label{yangmillsp}
\end{equation}
where the color current associated to each colored particle is 
\begin{equation} 
j_{\mu} ^a (x) = g \int d \tau \,   p_{\mu}(\tau)\,   Q^a (\tau) 
\, \delta^{(4)} ( x - y (\tau)) \ .
\label{punc}
\end{equation}

As a consequence of the definition (\ref{defcol}), the color fields
obey the non-Abelian Bianchi identity
\begin{equation}
[D_\nu \, ^*F^{\nu\mu}]^a(x) =  0 \ ,
\end{equation}
where $^*F^{\nu\mu\,a} = \frac12 \epsilon^{\nu\mu \rho \sigma} F_{\rho
\sigma}^a$.

The main difference between the equations of electromagnetism
(\ref{lorentz}) and the Wong equations (\ref{wongeq}),
 apart from their intrinsic non-Abelian structure, comes
from the fact that color charges precess in color space, and therefore
they are dynamical variables. Equation (\ref{wongeqc}) guarantees 
that the color current associated to each colored particle
is covariantly conserved
\begin{equation}
 \left(D_{\mu} j^{\mu} \right)_a (x) =  \partial_{\mu} j^{\mu}_a  (x)
+ g f_{abc} A_{\mu}^b (x) j^{\mu c}  (x) = 0 \ ,
\label{covcons}
\end{equation}
 therefore preserving the consistency of the theory. Notice that
(\ref{covcons}) is required as a compatibility condition, 
since in applying a covariant derivative 
$D_{\mu}$ to (\ref{yangmillsp}) the l.h.s of the equation should
vanish.

Let us formulate the statistical description of a plasma of colored
particles in their phase-space.
The usual $(x,p)$ phase-space is  enlarged to $(x,p,Q)$ by 
including  color degrees of freedom for colored particles. 
Physical constraints are enforced by inserting delta-functions in 
the phase-space volume element $dx\,dP\,dQ$. The momentum 
measure, chosen as in (\ref{measurep}),
guarantees positivity of the energy and on-shell evolution. The color 
charge measure enforces the conservation of the group invariants, 
{\it e.g.}, for $SU(3)$,
\begin{equation}
dQ = d^8 Q\,\, \delta(Q_{a}Q^{a} - q_{2})\,\, 
\delta(d_{abc}Q^{a}Q^{b}Q^{c} - q_{3}) \ , 
\label{measureq}
\end{equation}
where the constants $q_{2}$ and $q_{3}$ fix the values of the 
Casimirs and $d_{abc}$ are the totally symmetric group constants. 
The color charges which now span the phase-space are dependent 
variables. These can be formally related to a set of independent 
phase-space Darboux variables \cite{KLLM}. For the sake of simplicity, 
the standard color charges will be used in the remaining part of this Section.

The one-particle distribution function $f(x,p,Q)$ denotes the 
probability for finding the particle in the state $(x,p,Q)$.
In the collisionless case,  it evolves in 
time via a transport equation 
$ {{d f}\over{d \tau}} = 0$. Using the equations of motion~(\ref{wongeq}), 
it becomes the Boltzmann equation \cite{EH}
\begin{equation}
p^{\mu}\left[{{\partial}\over{\partial x^{\mu}}}
- g\, Q_{a}F^{a}_{\mu\nu} (x) {{\partial}\over{\partial p_{\nu}}}
- g\, f_{abc}\,A^{b}_{\mu} (x) Q^{c}{{\partial}\over{\partial Q_{a}}}
\right] f(x,p,Q) = 0 \ .
\label{boltzmann}
\end{equation}

A complete, self-consistent set of non-Abelian Vlasov equations for 
the distribution function and the mean color field is obtained by 
augmenting the Boltzmann equation with the Yang-Mills equations:
\begin{equation}
[D_\nu F^{\nu\mu}]^a(x) = J^{\mu\, a}(x) =  \sum_{\rm species}\
 \sum_{\rm helicities}\ 
j^{\mu\, a}(x)\ ,
\label{yangmills}
\end{equation}
where the  color current $j^{\mu\,a}(x)$
for each particle species is 
computed from the corresponding distribution function as
\begin{equation}
j^{\mu\,a} (x) = g\, \int dPdQ\ p^\mu Q^a f(x,p,Q) \ .
\label{cr5}
\end{equation}
Notice that if the particle's trajectory in phase-space would be known exactly,
then Eq. (\ref{cr5}) could be expressed as in Eq. (\ref{punc}). 
Furthermore, the color current (\ref{cr5}) is covariantly conserved,
as can be shown by using the 
Boltzmann equation \cite{KLLM}.  

 The Wong equations (\ref{wongeq}) 
are  invariant  under the finite gauge 
transformations (in matrix notation)
\begin{equation}
\label{gaugetrsf} 
\bar{x}^{\mu}=x^{\mu}\ , \qquad
\bar{p}^{\mu}= p^{\mu}\ , \qquad 
\bar{Q} =  U \,Q \,U^{-1}\ , \qquad
{\bar A}_\mu = U\,A_\mu \,U^{-1}-{1\over g}\,U\,
{\partial\over \partial x_\mu}\,U^{-1}\ ,
\end{equation} 
$\!\!$where $U=U(x)$ is a group element.

It can be shown \cite{KLLM} that the Boltzmann equation (\ref{boltzmann})
 is invariant under the above gauge transformation if
the distribution function behaves as a scalar
\begin{equation}
{\bar f}({\bar x},{\bar p},{\bar Q}) = f (x,p,Q) \ .
\end{equation}
To check this statement it is important to note that under a gauge 
transformation the derivatives appearing in the Boltzmann equation 
(\ref{boltzmann}) transform as \cite{KLLM}:
\begin{equation} 
{\partial\over\partial x^\mu}=
{\partial\over\partial\bar{x}^\mu}
- 2 ~{\rm Tr}~ \Biggl([\ ({\partial\over\partial {\bar x}^\mu}U) 
U^{-1}\ ,\  \bar{Q}\ ] 
{\partial\over\partial\bar{Q}}\Biggr) \ , \qquad
{\partial\over\partial p^\mu}=
{\partial\over\partial\bar{p}^\mu} \ , \qquad
{\partial\over\partial Q}=
U^{-1} {\partial\over\partial\bar{Q}}U
\label{eq:gauge2c}\ , 
\end{equation}
that is, they are not gauge invariant by themselves. Only
the specific combination of the spacial and color derivatives
that appears in (\ref{boltzmann}) is
gauge invariant.

 The color current 
(\ref{cr5}) transforms under (\ref{gaugetrsf}) as a gauge covariant vector: 
\begin{equation}
{\bar j}^{\mu}({\bar x})= g\,\int dP\,dQ\,p^\mu\,U\,Q\,
U^{-1}\,f(x,p,Q)=U\,j^{\mu}(x)\,U^{-1}\ .
\end{equation}
This is due to the gauge invariance of the phase-space measure and to the
transformation properties of $f$.

Let us consider now a gluon plasma, that is, a plasma of
particles  carrying a non-Abelian charge in the adjoint
representation, and which is initially at equilibrium. The system is 
disturbed by a weak color electromagnetic field, and one looks for the
response of the plasma. The distribution function can be expanded in
powers of $g$ as:
\begin{equation}
f=f^{(0)}+gf^{(1)}+...\ ,
\label{eq:2.2}
\end{equation}
where $f^{(0)}$ is, up to a normalization constant, the
Bose-Einstein equilibrium distribution function
\begin{equation}
n_B (p_0) = \frac{1} {e^{p_0/T} - 1} \ .
\end{equation}

 The Boltzmann equation (\ref{boltzmann}) for $f^{(1)}$ reduces to \cite{KLLM}
\begin{equation}
p^{\mu} \left({\partial\over\partial x^{\mu}}-g\, f^{abc}\, A_{\mu}^b (x)
Q_c {\partial\over\partial Q^a}\right) 
f^{(1)}(x,p,Q) = p^{\mu} Q_a F_{\mu 0}^a (x) {d \over d 
p_{0}} f^{(0)}(p_0)\ .
\label{eq:2.3}
\end{equation}
Notice that a complete linearization of the equation in $A_{\mu}^a$ 
would break the gauge invariance of the transport equation, which is
preserved in this approximation.
 But notice as well that this 
approximation tells us that $f^{(1)}$ also carries a $g$-dependence.

One can get the equation that the 
 color current  density $J^{\mu}_a (x,p)$ obeys by multiplying
(\ref{eq:2.3}) by $p^{\mu}$ and $Q_a$ and  then integrating over the color
charges. For gluons in the adjoint representation
\begin{equation}
\int dQ \, Q_a Q_b = N \, \delta_{ab} \ ,
\label{col}
\end{equation}
one finally gets, after summing over helicities,
\begin{equation}
[\,p \cdot D\,\, J^{\mu}(x,p)]^a =  2 \,g^ 2 \,N \, p^{\mu} p^{\nu} 
F_{\nu 0}^a (x) \, {d \over  d 
p_{0}} f^{(0)}(p_0) \ .
\end{equation}
Notice that only the color electric field enters in the r.h.s.
of the above equation. Thus, only the color electric field is
screened in the static situation.

The induced color current can be expressed in terms of the
parallel transporter $\Phi$ as \cite{Blaizot1}
\begin{equation}
\label{cmagcurr}
J^{\mu} _a (x) = m^2 _D  \int \frac{d \Omega_{\bf {\hat v}}}{4 \pi} \,
v^{\mu} \, \int^{\infty}_{0} du \, \Phi_{a b} (x, x- v u)\,  {\bf v} \cdot
{\bf E}^b (x - vu)
\, 
\end{equation}
where $m^2 _D = g^2 T^2 N/3$ is the Debye mass squared, and
$\Phi$ obeys the equation
\begin{equation}
\frac{\partial} {\partial u} \Phi_{a b} (x, x- v u) =  g\, \Phi_{a c} (x, x- v u) \, f_{c b d} 
\,v^{\mu} A_{\mu}^d (x- v u)
\label{partrans}
\end{equation}
with the initial boundary condition $\Phi_{a b} (x,x) = \delta_{a b}$.
The four vector $v^{\mu}$ is the velocity vector of the particles of the
plasma, and in the ultrarelativistic limit is light-like. Retarded boundary
conditions have also been implemented in Eq. (\ref{cmagcurr}).

Alternatively, in momentum space $J^{\mu} _a (k)$  may be expressed
as an infinite power series in the vector gauge field $A_{\mu} ^a (k)$ 
\cite{JacNa}.

In the static situation the color current simplifies drastically and may
be expressed as
$J_{\mu} ^a ({\bf x}) = - m^2 _D \, \delta_{\mu 0} A_0 ^a ({\bf x})$ \cite{JLL}.

The non-Abelian Kubo equations were first derived in Ref. \cite{JacNa}.
In the static limit they read 
\begin{eqnarray}
\left( {\bf D} \cdot {\bf E} \right)_a & = & J^{0} _a ({\bf x}) \ ,
\qquad  \left( {\bf D} \times  {\bf B} \right)_a + g f_{abc}\, A_0 ^b \, {\bf E} ^c = 0 \ , \\
\left( {\bf D} \cdot {\bf B} \right)_a  &= & 0 \ , \qquad
\qquad  \left( {\bf D} \times  {\bf E} \right)_a -  g f_{abc}\, A_0 ^b \, {\bf B}^c = 0 \ . 
\end{eqnarray}

The static non-Abelian Kubo equations have been studied in Ref. \cite{JLL}
and they describe the static screening of color electric fields. Furthermore,
the color electric screening effects described by these equations are 
reproduced in the context of resummed perturbative QCD
in the high temperature limit \cite{BP}.

\subsection{Static Screening of Color Magnetic Fields in the Plasma
of Non-Abelian Magnetic Charges}

In this subsection  the screening effects in a plasma of non-Abelian
magnetic charges are derived. Those particles are the natural
non-Abelian analogues of the
magnetic Dirac monopoles that were studied in Subsec. \ref{DAbDebye}.

As opposed to what happened in the Abelian theory, one cannot appeal
to duality symmetry arguments to describe the dynamics of non-Abelian
magnetic charges. It is a well-known fact that a pure Yang-Mills theory
without matter is not symmetric under the interchange of color electric and
magnetic degrees of freedom \cite{Deser}. This may be understood as follows.
In the absence of matter, the Yang-Mills equation and the non-Abelian
Bianchi identity both involve a covariant derivative, but not a ``dual'' 
covariant derivative. Thus, in terms of the dual field $^*F^{\mu \nu}_a$
the Yang-Mills equation cannot be interpreted as the non-Abelian
Bianchi identity for $^*F^{\mu \nu}_a$. In general, it is not possible to
find a dual vector gauge field ${\tilde A}^a_{\mu}$ which is related to
$^*F^{\mu \nu}_a$ as $A^a_{\mu}$ is related to
$F^{\mu \nu}_a$  in (\ref{defcol}).
 It has been realized in the literature that the dual of a 
Yang-Mills theory is a Freedman-Towsend  like theory
\cite{FT}, \cite{Lozano}. The fundamental
object of that theory is an antisymmetric two index tensor,  
${\tilde F}^{\mu \nu}_a$.
One can define in that theory a vector gauge field potential $V_{\mu}^a$ which
acts as a parallel transport for the phases of the charged particles,
 but which is not related to   ${\tilde F}^{\mu \nu}_a$
as in a Yang-Mills theory.

The equations of motion of a particle of mass $m$, carrying a non-Abelian
magnetic charge of $SU(N)$ ${\tilde Q}^a$, with  $a=1,...,N^2-1$,  and
transversing a worldline $y^{\mu} (\tau)$ have been derived in
Ref. \cite{HongMo}. The derivation was made there using a variational principle
in loop space.  The equations  written in terms of the 
color gauge fields are
\begin{mathletters}
\label{dwongeq}
\begin{eqnarray}
m\, {{d  y^{\mu}(\tau) }\over{d \tau}} & = &  p^{\mu} (\tau) 
\ , \label{dwongeqa} \\[2mm]
m\, {{d  p^{\mu} (\tau)}\over{d \tau}} & = &
 {\tilde g} \, {\tilde Q}^a (\tau)\, ^*F^{\mu\nu}_{a} ( y(\tau)) \,  p_{\nu} (\tau)
\ ,\label{dwongeqb}\\[2mm]
m\, {{d {\tilde Q}^{a} (\tau) }\over{d \tau}} & = & - g\, 
f^{abc} \,  p^{\mu} (\tau) \,
A^{b}_{\mu} ( y(\tau)) \, {\tilde Q}^{c} (\tau)\ . \label{dwongeqc}
\end{eqnarray}
\end{mathletters}
$\!\!$These are augmented with the field equations
\begin{eqnarray}
[D_{\nu} F^{\nu\mu}]_a (x) & = & 0 \ , \\[2mm]
[D_{\nu} \, ^*F^{\nu\mu}]_a (x) & = & {\tilde J}^{\mu}_a (x) =  \sum_{\rm species} \,
 \sum_{\rm helicities} {\tilde j}^{\mu}_a (x)\ ,
\label{dyangmillsp}
\end{eqnarray}
where  
\begin{equation} 
{\tilde j}_{\mu} ^a (x) = {\tilde g} \int d \tau \,   p_{\mu}(\tau)\,  {\tilde Q}^a (\tau) 
\, \delta^{(4)} ( x - y (\tau)) \ .
\label{cpunc}
\end{equation}

The above current is covariantly conserved $[D_{\mu} {\tilde J}^{\mu}] _a(x) =0$,
which is required by consistency, as may be realized in applying a covariant
derivative to (\ref{dyangmillsp}).

Notice that a new coupling constant ${\tilde g}$ appears in the equations
as coupling between the color electromagnetic fields and the non-Abelian
magnetic charges. In principle, this is a new variable
in the system. However,  there is a Dirac
quantization condition relating $g$ and ${\tilde g}$. We will come back
to this point later on. 

The above equations are the dual of the Wong equations written in 
 Subsec. \ref{NAbDebye}. Notice that they are not symmetric under the 
interchange of color magnetic and electric degrees of freedom, since the same
covariant derivative $D_{\mu}$ appears in both of them. Some caution
should be taken at this stage. Exactly as it happened in the Abelian situation, 
in the absence of the non-Abelian Bianchi identity one cannot define
globally a vector gauge field $A_{\mu}^a$ obeying the field equations. 
It could be defined locally, outside the monopole worldline. It is possible to
define different $A_{\mu}^a$ in different space-time regions, the different
solutions or patches
 being related in their common domain of definition by a gauge
transformation.

Let us stress that the non-Abelian electric and magnetic charges
$Q_a (\tau)$ and ${\tilde Q}_a (\tau)$ live in the same group manifold, and obey
the same kind of dynamical evolution. Both transform under gauge 
transformations in the same way.

With the above classical equations of motion it is possible to
derive the dual non-Abelian Vlasov equations. 
They read
\begin{mathletters}
\begin{eqnarray}
& p^{\mu}&\left[{{\partial}\over{\partial x^{\mu}}}
-{\tilde g} \,{\tilde Q}_{a} \, ^*F^{a}_{\mu\nu} (x) {{\partial}\over{\partial p_{\nu}}}
- g\, f_{abc}\,A^{b}_{\mu} (x){\tilde  Q}^{c}{{\partial}\over{\partial{\tilde  Q}_{a}}}
\right]{\tilde  f}(x,p,{\tilde  Q}) =   0 \ , \\[2mm]
\label{dboltzmann}
& [D_\nu & F^{\nu\mu}]^a(x) =  0 \ , \\ [2mm]
&[D_\nu  &\, ^*F^{\nu\mu}]^a(x) =
{\tilde J}^{\mu\, a}(x) =  \sum_{\rm species}\
 \sum_{\rm helicities}\ 
{\tilde j}^{\mu\, a}(x)\ ,
\label{dKmills}
\end{eqnarray}
\end{mathletters}
where 
\begin{equation}
{\tilde j}^{\mu\,a} (x) ={\tilde g} \, \int dPd{\tilde Q}\ p^\mu {\tilde Q}^a 
{\tilde f} (x,p,{\tilde Q}) \ .
\label{dcr5}
\end{equation}
The color magnetic charge measure $d{\tilde Q}$ is defined as in Eq. (\ref{measureq}).

The gauge transformation properties of the 
above equations will not be discussed here,
as they turn out to be exactly the same as their dual partners. 

Let us derive now the response of the non-Abelian magnetic
 plasma to a weak field disturbance.
Let us suppose that the neutral plasma of non-Abelian magnetic charges 
in the adjoint representation is
initially at equilibrium. Then one can look for a solution of the form
\begin{equation}
{\tilde f} = {\tilde f}^{(0)}+{\tilde g}{\tilde  f}^{(1)}+...\ ,
\label{deq:2.2}
\end{equation}
where  ${\tilde  f}^{(0)}$ is, up to a normalization constant,
 the Bose-Einstein equilibrium distribution function.
Then ${\tilde  f}^{(1)}$ obeys the equation
\begin{equation}
p^{\mu} \left({\partial\over\partial x^{\mu}}-g\, f^{abc}\, A_{\mu}^b (x)
{\tilde Q}_c {\partial\over\partial{\tilde  Q}^a}\right) 
{\tilde f}^{(1)}(x,p,{\tilde Q}) = p^{\mu} {\tilde Q}_a \, ^* F_{\mu 0}^a (x) {d \over d 
p_{0}}{\tilde f}^{(0)}(p_0)\ .
\label{deq:2.3}
\end{equation}

 A weak coupling expansion in ${\tilde g}$,
the natural coupling constant of the problem, has been performed above.
To preserve the gauge symmetry of the above Boltzmann equation, 
the term linear  in $A_\mu ^a$ in the l.h.s. of (\ref{deq:2.3}) has to be kept.
This tells us that ${\tilde f}^{(1)}$  has  a  dependence 
on the coupling constant $g$. The two constants $g$ and ${\tilde g}$ are
related by a Dirac quantization condition that  will be discussed later on.

From  Eq. (\ref{deq:2.3}) one can derive the static screening of
color magnetic fields. In order to do that, one should follow the same procedure 
as  in the previous subsections. Notice that now {\it only} the color
magnetic field enters in the r.h.s of (\ref{deq:2.3}).

The equation obeyed by the current density is obtained after integrating
over the magnetic charges ${\tilde Q}_a$, and summing over helicities, 
and it reads
\begin{equation}
[\,p \cdot D\, {\tilde J}^{\mu}(x,p)]^a = 2 \, {\tilde g}^ 2 \,  N\, p^{\mu} p^{\nu} 
\, ^*F_{\nu 0}^a (x) \, {d \over  d 
p_{0}}{\tilde  f}^{(0)}(p_0) \ .
\end{equation}

The solution of the above equation may be written as 
\begin{equation}
{\tilde J}^{\mu} _a (x) = {\tilde m}^2 _D  \int \frac{d \Omega_{\bf {\hat v}}}{4 \pi} \,
v^{\mu} \, \int^{\infty}_{0} du \, \Phi_{a b} (x, x- v u)\,  {\bf v} \cdot
{\bf B}^b (x - vu) \ ,
\, 
\end{equation}
where $\Phi$ is the parallel transporter, which obeys also
Eq. (\ref{partrans}).
Here ${\tilde m}_D ^2 ={\tilde g}^2 T^2  \,N/3 $ is the magnetic Debye mass
squarred. 

The dual Non-Abelian Kubo equation read in the static limit, 
therefore 
\begin{eqnarray}
\left( {\bf D} \cdot {\bf E} \right)_a & = & 0 \ , \qquad \qquad 
 \left( {\bf D} \times  {\bf B} \right)_a + g f_{abc}\, A_0 ^b \, {\bf E} ^c = 0 \ , \\
\left( {\bf D} \cdot {\bf B} \right)_a  &= & {\tilde J}^{0} _a ({\bf x})  \ , \qquad
\left( {\bf D} \times  {\bf E} \right)_a -  g f_{abc}\, A_0 ^b \, {\bf B}^c = 0 \ ,
\end{eqnarray}
and  they describe the static screening of color magnetic fields.

There is a Dirac quantization condition relating the two coupling constant
$g$ and ${\tilde g}$, which depends on the Lie group under consideration. 
For the case  considered here, where  all 
the matter is in the adjoint representation, then the gauge field theory
which is based
upon the Lie algebra $su (N)$ has as a global Lie group $SU (N)/Z_N$ and
not  $SU (N)$.  Here $Z_N$ is the $N$-element finite group consisting in the
$N$-th roots of unity, that is, the integral powers of $\exp{(2 \pi i/N)}$. 
This is due to the fact that two  $SU (N)$ matrices that differ only by a factor
belonging to $Z_N$ will be represented by the same matrix in the adjoint
representation. For a  $SU (N)/Z_N$ theory the Dirac quantization condition
reads \cite{Coleman}, \cite{HongMo}, \cite{GNO}
\begin{equation}
g \, {\tilde g} = \frac{2 \pi \,  n} { N} \qquad  n \, \epsilon  \, Z \ .   
\end{equation}
Therefore, the dual Debye mass is written in terms of $g$ as
\begin{equation}
{\tilde m}_D = \frac{T}{ g \sqrt{3 N}}\, 2 \pi \, n
\end{equation}

Therefore, for small values of $N$ 
 the static magnetic screening effects could not have been
reproduced by using  perturbation theory in $g$, as then
the expansion of Eq. (\ref{deq:2.2})
would  not hold. However, notice that for $N \rightarrow \infty$, both
$g$ and $\tilde{g}$ can both be small, and therefore perturbative expansions in
$g$ and $\tilde{g}$ can both be valid.

\section{ STATIC MAGNETIC MONOPOLE FIELDS}

In this section  exact solutions to the dual Kubo equations are found. Those
solutions describe screened magnetic fields in the plasma of magnetic charges.
Some known results on how to construct magnetic monopole fields
in the vacuum for the Abelian and non-Abelian theories are first reviewed.

\subsection{Monopole Fields in the Vacuum}
\label{Vacsmon}

The magnetic field created by a magnetic charge ${\tilde e}$ which
is at rest at the origin of coordinates is given by
\begin{equation}
{\bf B} = \frac{{\tilde e}}{4 \pi} \, \frac{\bf r}{r^3} \ .
\label{magmon}
\end{equation}
As already explained in Subsec. \ref{DAbDebye}, it is not possible to construct a global vector
gauge field $A_{\mu} (x)$ which generates the above magnetic field. However,
it is possible to find a local vector gauge field which is defined everywhere
except on a ``Dirac string". That was the original construction of magnetic
monopole fields due to Dirac.
Wu and Yang \cite{WuYang2} showed that it is also possible to construct vector gauge
fields without references to Dirac strings as follows.  Let us denote by $R$ 
the space region surrounding the magnetic monopole. Dividing 
 $R$ into two regions, $R_+$ and $R_-$ defined as  (in spherical
coordinates $(r, \theta, \phi)$),
\begin{eqnarray}
R_+ : & \qquad 0 \leq \theta < \pi/2 + \delta , \qquad r > 0, \qquad 
0 \leq \phi < 2 \pi &  \ , \\
R_- : & \qquad   \pi/2 - \delta < \theta  \leq \pi , \qquad r > 0, \qquad 
0 \leq \phi < 2 \pi & \ ,
\end{eqnarray}
with an overlap region $\pi/2 - \delta < \theta  <  \pi/2 + \delta$,
where $0 < \delta \leq \pi/2$. In each
region one may take \cite{WuYang2}
\begin{mathletters}
\label{WYmnf}
\begin{eqnarray}
{\bf A}_+ & = & \frac{{\tilde e}}{4 \pi r} \frac{(1 -\cos{\theta})}{\sin{\theta}}\, 
{\hat {\bf  e}} _{\phi} \ , \\ 
{\bf A}_- & = & - \frac{{\tilde e}}{4 \pi r} \frac{(1 +\cos{\theta})}{\sin{\theta}}\, 
{\hat {\bf e }}_{\phi} \ . 
\end{eqnarray}
\end{mathletters}
These two vector gauge potentials reproduce the magnetic field 
(\ref{magmon}) in their respective domain of definition. Furthermore,
in the overlap region the two vector gauge fields are related by a 
gauge transformation
\begin{equation}
{\bf A}_+ - {\bf A}_- = \frac{{\tilde e}}{2 \pi r} \, {\hat {\bf e}}_ {\phi} =
 {\bf \nabla} \left(\frac{ {\tilde e} \phi}{2 \pi} \right) \ .
\end{equation}

In a quantum description, this gauge transformation implies  a change
in the wavefunction $\psi$ of a 
particle , $\psi \rightarrow  \exp{ ( - i e {\tilde e} \phi / 2\pi) \psi}$.
After requiring the wavefunction be single-valued, one arrives at the
Dirac quantization condition $e {\tilde e} = 2 \pi  n$.

One can easily generalize the previous results to the non-Abelian case.
Let us consider first for definiteness the $SU(2)$ Lie group and take as its
infinitesimal generators $t_a = -\frac {i}{2} \sigma_a$, where $\sigma_a$ are
the Pauli matrices. A non-Abelian magnetic monopole field may be constructed
from the Abelian one by taking the same value (\ref{magmon}) in a particular
direction in color space. For example, 
\begin{equation}
{\bf B}_3 = \frac{{\tilde g}}{4 \pi} \, \frac{\bf r}{r^3} \ , \qquad 
{\bf B}_1 = {\bf B}_2 = 0 \ .
\label{SU2magmon}
\end{equation}
The vector gauge fields which reproduce the above color magnetic
field are obtained from taking in the $a =3$ direction in color space
the Abelian vector gauge fields (\ref{WYmnf}), while $A^{\mu} _1 = A^{\mu} _2
=0$.

In general, for a simple Lie group, the corresponding magnetic monopole
fields can be constructed from their Abelian counterparts, just by multiplying
the Abelian fields by a constant matrix ${\tilde  Q}$ living in a
specific representation of the Lie group. Those are solutions  to the Yang-Mills
equations in the presence of a static point magnetic source 
 ${\tilde J}_{\mu}  = \delta_{\mu 0}\, {\tilde g}\, {\tilde Q}  \delta^{(3)} ({\bf r})$.

Due to the Abelian character of these non-Abelian monopoles,
the  Dirac quantization 
condition can be easily worked out, and it reads
\begin{equation}
\exp{\left( 2 \pi  \frac{g {\tilde g} {\tilde Q}} {2 \pi}\right)} = 1  \ .
\label{YMDQC}
\end{equation}
The above quantization condition is sensitive to the global structure of
the Lie group under consideration, and it could be different for different
Lie groups sharing the same Lie algebra. 
For example, for $SU(2)$ and $SO(3)$, Eq. (\ref{YMDQC}) has different
implications \cite{WuYang2}. 

There is a dynamical \cite{GNO}, as well as topological classification
\cite{Coleman} of these non-Abelian monopoles. The topological classification
of the monopoles associates each class of monopole to each different
element of the first homotopy group $\pi_1 (G)$
of the Lie group $G$ under consideration. A stability analysis of the
non-Abelian magnetic monopole fields was performed in \cite{BraNe},
finding that there
is one stable non-Abelian monopole field for each topological class.

\subsection{Static Magnetic Monopole Fields in the Hot Plasmas}
\label{Hotsmon}

In this subsection  some exact solutions to the static dual 
Abelian and non-Abelian Kubo field equations are presented.
In order to find  those
solutions, rather than solving the corresponding differential equations, we
will take profit of the following facts. First, in the Abelian situation, the solutions
to the ``magnetic" Kubo equations can be found from those of the ``electric" ones,
 just by making use of the duality symmetry of electromagnetism.
Second, solutions to the non-Abelian Kubo equations can always be
constructed from the Abelian ones, as we have already
explained.  Using these facts, one can easily
construct solutions to the dual non-Abelian Kubo equations
which have an Abelian character.

Let us first recall how to find exact solutions to the static Kubo equations
 (\ref{Kuboabel}). Those can be easily solved in terms of the electric  potential
$A_0$, with  ${\bf E} = - {\bf \nabla} A_0$. Then, the equation obeyed by
$A_0$ is
\begin{equation}
\left( {\bf \nabla}^2 - m_D ^2 \right) A_0 ( {\bf x}) = 0 \ ,
\label{spherbess}
\end{equation}
With   spherical symmetry, and in spherical coordinates $(r, \theta, \phi)$,
 Eq. (\ref{spherbess})
can be easily separated. The angular part of the solution is given in terms
of spherical harmonics, and the radial part is expressed in terms of modified
spherical Bessel functions. For the first harmonic, or equivalently, the monopole
term in a multipole expansion, and discarding the solution growing exponentially,
one then finds 
\begin{equation}
 A_0  = a_0 \frac{e^{-m_D r}} {r}  \ , \qquad A_i = 0\ .
\end{equation}
which corresponds to a  screened radial electric field.

The non-Abelian Kubo equations have been studied in the literature
using different {\it Ans\"{a}tze} \cite{Kubo}.  The static case  for the $SU(2)$ group
with spherical symmetry was considered 
in Ref. \cite{JLL}.
 Two particular solutions were found there. The first one corresponds to the
Yang-Mills vacuum. The second one  can actually be constructed 
 by taking
the Abelian solution in a specific direction of isospin space. Thus 
\begin{equation}
 A_0 ^a  =  {\hat r}^a a_0 \frac{e^{-m_D r}} {r} \ ,  \qquad A_i ^a = 0 \ ,
\label{solJLL}
\end{equation}
which describes a screened color electric field. 
Pure non-Abelian solutions were also studied in  Ref. \cite{JLL}, but it was shown
there that they tend asymptotically to either the Yang-Mills vacuum or to 
(\ref{solJLL}).

Let us discuss now the magnetic Kubo equations.
In the Abelian case, it is possible to find solutions
to the Eq. (\ref{DKuboabel}), just by writing the magnetic field in terms of a ``magnetic"
potential, ${\bf B} = - {\bf \nabla} {\tilde A_0}$.  This is possible since in 
this case ${\bf \nabla} \times {\bf B} = 0$, as we have already mentioned.
Therefore, in terms of 
${\tilde A_0}$, the field equation becomes
\begin{equation}
\left( {\bf \nabla}^2 - {\tilde m}_D ^2 \right) {\tilde A}_0 ( {\bf x}) = 0 \ ,
\end{equation} 
which is exactly the same equation as  (\ref{spherbess}), and therefore,
has the same solutions. In particular, with spherical symmetry,
 the monopole solution is
\begin{equation}
{\tilde  A}_0  = {\tilde a}_0 \frac{e^{-{\tilde m}_D r}} {r} \ ,   \qquad {\tilde  A}_i = 0 \ ,
\end{equation}
which gives  the screened radial magnetic field
\begin{equation}
{\bf B} =  - {\tilde a}_0 \frac{\bf r} {r^3} \left(1 + {\tilde m}_D r \right)
 e^{-{\tilde m}_D r}
\end{equation}

Let us  consider now the non-Abelian case, for the group $SO(3) \equiv SU(2)/Z_2$.
In this case it is possible to construct easily solutions 
from the Abelian ones.  It is enough to consider vanishing  vector gauge field 
configurations in all except one specific direction in color space.
Then the dual non-Abelian Kubo equations reduce to the dual Abelian ones. For
example, one solution is given by taking in the third direction of isospin space
\begin{equation}
{\bf B}_3 = - {\tilde a}_0 \frac{\bf r} {r^3} \left(1 + {\tilde m}_D r \right)
 e^{-{\tilde m}_D r} \ , \qquad {\bf B}_1 = {\bf B}_2 = 0 \ ,
\end{equation}
which describes a screened color magnetic field.

\section{DISCUSSION}
\label{conclu}

The purpose of this article has been finding  a classical
mechanism which generates  the thermal screening of magnetic fields.
It has been shown that at finite $T$, and in the presence
of magnetic charges, magnetic fields are screened. This effect can be 
easily understood in the Abelian theory. The duality symmetry of
electromagnetism allows to derive the magnetic screening from 
the electric one, without further complications.  A non-Abelian
theory is not symmetric under a duality transformation, but still
it is possible to show that also color magnetic fields can be screened.
In both cases, the effects of magnetic screening 
are proportional to the inverse of the gauge coupling
constant.

The question which remains to be answered is whether the polarization
effects that have been described in this article could be relevant in realistic
quantum gauge theories, such as QED or QCD. The question can be actually 
reduced to answer if there are magnetic monopoles in those theories.  

It seems obvious that QED does not possess magnetic monopoles.
Thus, it is not expected that the mechanism described in this article
takes place in QED.
A pure Abelian gauge theory does not suffer from IR
problems. The IR divergences of QED that arise in the magnetic sector of the
theory at finite $T$ can actually be
cured in the same way than at zero $T$ \cite{BI2}.

In a pure Yang-Mills theory 't Hooft \cite{`t Hooft} showed 
that in the maximal Abelian gauge that theory has  magnetic monopoles.
Actually this  occurs in several different Abelian projections of the
non-Abelian theory.
The idea of describing the QCD ground state as a condensate of magnetic
monopoles has deserved much attention \cite{Mandest}, \cite{Poly},
since then one could
understand the confinement of QCD as a dual Meissner effect.
Although these ideas are very appealing, there is not yet a complete
 gauge independent analytical proof
of this confinement scenario, except for the case of some supersymmetric
non-Abelian models \cite{Seiberg}. However, lattice computations have shown  the
magnetic monopole dominance for the string tension of QCD at zero temperature
\cite{lattice}.

 A natural expectation is that if magnetic monopoles explain the confinement 
of QCD because they condensate at zero temperature, they should also play an
important role in explaining the magnetic screening at finite temperature.
I have shown that this screening would occur by considering
a plasma of magnetic charges.
Let me stress that the real situation corresponding to the real QCD plasma
would be much more complicated  than the simple models that  I have described, as then both electric and
magnetic polarization phenomena should occur simultaneously. There is also a 
mismatch between the magnetic mass
of order $T/g$ that has been derived here, and an expected one 
of order $g^2 T$.
Our result has been obtained on the assumption that the equilibrium
density of monopoles is of order $T^3$. However
this does not have to be so necessarily.
 To recover the expected magnetic mass
of order $g^2 T$, the equilibrium density of magnetic monopoles should be
of order $(g^2 T)^3$. 
This would naturally imply that magnetic monopoles are not elementary particles,
but that they should be dynamically generated.

  Lattice computations could check whether the mechanism
of  magnetic screening that I have described takes place or
not in a pure Yang-Mills theory at finite temperature.

\vskip 1cm

{\bf Acknowledgments:}

I want to thank J.~Ambj{\o}rn, J.~I.~Latorre, J.~Paris and J.~Roca
for  useful conversations, and to Y.~Lozano for instructive e-mail discussions on 
S-duality.

This work has been supported  by funds provided by the
CICYT contract AEN95-0590, and by the CIRIT contact GRQ93-1047.

\end{document}